\documentclass[aps,showpacs,floats,letterpaper,floatfix,groupedaddress,eqsecnum]{revtex4}
\usepackage{bm}
\usepackage{epsfig,amsopn}
\usepackage{graphicx}
\usepackage{amsmath,amssymb}
\usepackage{natbib}
\newcommand\bea{\begin{eqnarray}}
\newcommand\eea{\end{eqnarray}}
\newcommand\beq{\begin{equation}}
\newcommand\eeq{\end{equation}}

\newcommand{\bib}{\bibitem}

\def\nn{\nonumber}
\def\dg{\dagger}
\def\f{\frac}
\def\la{\langle}
\def\ra{\rangle}

\def\b{\beta}

\def\e{\epsilon}
\def\g{\gamma}

\def\ua{\uparrow}
\def\da{\downarrow}
\def\bk{{\bf k}}

\def\bl{{\bf l}}
\def\b0{{\bf 0}}

\begin{document}

\title{Two mesoscopic models of two interacting electrons}

\author{Dibyendu Roy$^\dagger$} 
\affiliation{$^{*}$Raman Research Institute, Bangalore 560080, India
}

\begin{abstract}
We study two simple mesoscopic models of interacting two electrons; first one consists of two quantum coherent parallel conductors with long-range Coulomb interaction in some localized region and the other is of an interacting quantum dot (QD) side-coupled to a noninteracting quantum wire. We evaluate exact two-particle scattering matrix as well as two-particle current which are relevant in a two-particle scattering experiment in these models. Finally we show that the on-site repulsive interaction in the QD filters out the spin-singlet two-electron state from the mixed  two-electron input states in the side-coupled QD model.     
\end{abstract}

\vspace{0.0cm}
\date{\today}

\maketitle

\section{Introduction}

The Landauer-B{\"u}ttiker (LB) scattering approach is the cornerstone in the study of quantum transport in noninteracting mesoscopic systems \cite{Landauer75, Buttiker90}. One can make an one-to-one connection between the Lippman-Schwinger (LS) scattering theory and the LB approach \cite{Mello04}. There are also several  theoretical approaches to incorporate Coulomb interaction between electrons to investigate the transport phenomena in interacting models \cite{Glazman88, Averin90, Wingreen94, Kamenev96, Gurvitz97}. But most of these techniques are either perturbative in the interaction/tunneling  strength or valid only in the linear response regime. Thus a full-fledged quantum transport method to study the interplay between the strong interaction and the nonequilibrium behavior is on-demand. One way to tackle this problem is to employ the time-independent elastic LS scattering theory and find an exact many-body scattering eigenstate of the open interacting system. The basic assumption here again as the original LB scattering approach is that all the dissipation is considered to occur only in the reservoirs connected to the mesoscopic sample. Recently there have been some studies \cite{Mehta06, Goorden07, Dhar08, Roy09, Lebedev08, Nishino09} along this direction. A model of two quantum coherent conductors interacting weakly via a long range Coulomb force locally in some region has been studied in \cite{Goorden07}. Both the LS and the LB approaches have been employed and the two-particle scattering matrix is expressed in terms of the scattering matrices of the noninteracting conductors. The results in \cite {Goorden07} are perturbative in the interaction strength. Here we study a similar model and show that it is possible to find $exactly$ the two-particle scattering matrix as well as the two-particle current change due to the interaction in this model. Later we investigate another interacting open quantum impurity model; an interacting quantum dot (QD) is side-coupled to a noninteracting quantum wire. We show that the on-site interaction in the QD filters out the two-particle spin-singlet state from the mixed two-particle input states. We here apply the technique developed in Ref.\cite{Dhar08, Roy09} based on the LS scattering theory. It has been shown in \cite{Dhar08, Roy09} that one can find an exact two-particle scattering state for certain open quantum impurity models. A many-particle scattering state has been found in \cite{Dhar08, Roy09} within a two-particle scattering approximation. Physically in a real two-particle scattering experiment \cite{J08} one considers two wavepackets representing two electrons. 
\section{Scattering of electrons between two interacting conductors:}
\begin{figure}
\begin{center}
\includegraphics[width=7cm]{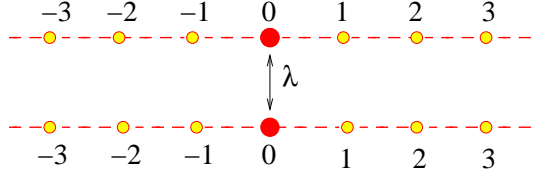}
\end{center}
\caption{(color online) A schematic description of the two coherent parallel conductors in the presence of finite on-site energy at the dot sites.}
\label{mag}
\end{figure}

We consider two quantum dots capacitively coupled via interaction $\mathcal{V}$. Both the dots are connected to two noninteracting leads modeled by one-dimensional tight-binding Hamiltonian, $\mathcal{H}^{\alpha}_L$.  Electron moves from one lead to other through the dot and interacts with electron of the other dot only at the dot sites.  But there is no exchange of electrons between the dots. This is a lattice version of the model studied in \cite{Goorden07}. We can better think of the model as of two separate parallel mesoscopic conductors (labelled by $I$ and $II$) in proximity of each other and single electron in each conductor (see Fig.\ref{mag}). Electrons in the conductors interact only in some localized region. For simplicity  we consider here spinless electrons. The Hamilton
\bea 
\mathcal{H} ~&=&~\mathcal{H}^I ~+~\mathcal{H}^{II} ~+~\mathcal{V}, ~~~{\rm where}~~\mathcal{H}^{\alpha}=\mathcal{H}^{\alpha}_L+\mathcal{H}^{\alpha}_D+\mathcal{V}^{\alpha}~~~{\rm with} \label{ham} \\
\mathcal{H}^{\alpha}_L&=&- \sum_{l=-\infty}^{\infty} \hspace{-0.15cm}' ~~(c_{\alpha,l}^\dg c_{\alpha,l+1} + c_{\alpha,l+1}^\dg c_{\alpha,l} )~,~ \mathcal{H}^{\alpha}_D= \epsilon_\alpha c_{\alpha,0}^\dg c_{\alpha,0}~,\nn \\
\mathcal{V}^\alpha&=&- \g^{\alpha} (c^\dg_{\alpha,-1}c_{\alpha,0}+c_{\alpha,0}^\dg c_{\alpha,-1}+c^\dg_{\alpha,0} c_{\alpha,1}+c_{\alpha,1}^\dg c_{\alpha,0}),~~{\rm and}~~\mathcal{V}=\lambda c_{I,0}^\dg c_{I,0}c_{II,0}^\dg c_{II,0}~.\nn
\eea
Above $\sum'$ implies omission of $l=-1,0$ from the summation and $\alpha=I,II$. Here $c_{\alpha,l}~(c_{\alpha,l}^\dg)$ is the electron annihilation (creation) operator in the $\alpha$th conductor. $\epsilon_{\alpha}$ is the on-site energy on the $\alpha$th QD and $\lambda$ is the strength of electrostatic Coulomb interaction between electrons in the two QDs. Also we set the hopping amplitude between the sites on both the conductors to unity. The double QDs with a purely capacitive interdot interaction can be labeled by a pseudospin index for the two dots and thus can be considered as a realization of the Anderson impurity model. One expects that electron transport through the QDs, that are weakly tunnel coupled to their leads, is dominated by the interdot interaction at low temperatures and this leads to Coulomb blockade \cite{Glazman88}. Recently H{\"u}bel {et al.} have shown that the interdot Coulomb blockade can be overcome by correlated tunneling when tunnel coupling to the leads is increased \cite{Hubel08}. We have here complete freedom within our approach to tune the tunnel junctions as well as the values of the dot energies and interdot interaction. Therefore we are able to study all regimes of the parameter space in our work.
\subsection{Scattering states}
We find exactly all the two-electron energy eigenstates for this model. First we need to calculate two-particle eigenstates of the noninteracting Hamiltonian $\mathcal{H}_0=\mathcal{H}^I+\mathcal{H}^{II}$ for $\lambda=0$. One-electron eigenstates $\phi^{\alpha}_k (l)$ $(=\la l| \phi^{\alpha}_k \ra)$ in a single conductor of Hamiltonian $\mathcal{H}^{\alpha}$ (with $\alpha=I,II$) can be found by solving the single electron Schr{\"o}dinger equation.
For an electron incoming from the left $0 < k < \pi$, the complete wave function is given by
\bea \phi^{\alpha}_k (l) &=& e^{ikl} ~+~ r^{\alpha}_k e^{-ikl} ~~{\rm for}~~ l \le -1, \nn \\
&=& (1+r^{\alpha}_k)/\g^{\alpha} ~{\rm for}~ l = 0, \nn \\
&=& t^{\alpha}_k e^{ikl} ~~{\rm for}~~ l \ge 1, \eea
with the following transmission (reflection) amplitudes $t^{\alpha}_k ~(r^{\alpha}_k)$,
\bea
t^{\alpha}_k=\f{2i{\g^{\alpha^2}}\sin k}{2{\g^{\alpha^2}}e^{ik}-\epsilon_{\alpha}-2\cos k},~{\rm and}~~ r^{\alpha}_k=t^{\alpha}_k-1~.\nn
\eea
Similarly one can find the single particle scattering state for a particle incident from the right.

We form a two-particle incoming state $\phi_{\bk} (\bl)$ of two electrons with one in each conductor. $\phi_{\bk} (\bl)=\phi^I_{k_1} (l_1)\phi^{II}_{k_2} (l_2) $, with $\bk=(k_1,k_2)$ and $\bl=(l_1,l_2)$. As electrons in the two conductors are distinguishable, we need not to anti-symmetrize the two-electron wave function. The energy of this state is $E_{\bk}=E_{k_1}+E_{k_2}$. 
A scattering eigenstate $|\psi_{\bk}\ra$ of $\mathcal{H}$ with energy $E_{\bk}$ is related to a state $|\phi_{\bk}\ra~(=|\phi_{k_1}^I\ra|\phi_{k_2}^{II}\ra)$ of $\mathcal{H}_0$ through the Lippman-Schwinger equation \cite{Mello04}
\bea |\psi^{\pm}_{\bk}\ra &=& |\phi_{\bk}\ra ~+~ G_0^{\pm}(E_{\bk}) \mathcal{V} |\psi^{\pm}_{\bk}\ra, \label{lipp}\\
{\rm where} ~~G_0^{\pm}(E_{\bk}) &=& \f{1}{E_{\bk}- \mathcal{H}_0 \pm i \e}. \nn \eea 
As usual, the indices $(\pm)$ indicate outgoing wave $(+)$ or incoming wave $(-)$ boundary conditions. Now in the two-electron sector, with the position basis $|\bl \ra$ and  an
incident state $<\bl|\phi \ra=\phi_\bk(\bl)$, Eq.(\ref{lipp}) gives
\bea \psi^+_\bk(\bl) &=& \phi_\bk(\bl) ~+~ \lambda K_{E_\bk}(\bl)~ \psi^+_\bk (\b0), 
\label{scatt} \\
{\rm where}~~ K_{E_\bk}(\bl) &=& <\bl| G_0^+(E_\bk) |\b0 \ra, \nn \eea
and $\b0 \equiv (0,0)$. We can determine $\psi^+_\bk(\b0)$ using Eq. 
(\ref{scatt}), 
\beq \psi^+_\bk(\b0) ~=~ \f{\phi_\bk(\b0)}{1-\lambda K_{E_\bk}(\b0)}. \label{psi0} \eeq
The two-electron scattering eigenstate is completely given by Eqs. 
(\ref{scatt}-\ref{psi0}). As it has been shown in \cite{Dhar08, Roy09} that after scattering from the interaction the total momentum of the scatterred particles is not conserved though the total energy remains same. The momenta $(k'_1,k'_2)$ of the scatterred particles are related to the incident momenta $(k_1,k_2)$ through, $\cos k_1+\cos k_2=\cos k'_1+\cos k'_2$. 
The matrix elements $K_{E_\bk}(\bl)$ are known explicitly and are given by
\bea K_{E_\bk}(\bl) = ~ \int_{-\pi}^\pi \int_{-\pi}^\pi \f{dq_1 dq_2}{(2 
\pi)^2} ~\f{1}{E_\bk - E_{\bf q} +i \e} \phi^I_{q_1} (l_1) \phi^{II}_{q_2} (l_2) \phi^{I*}_{q_1} (0) \phi^{II*}_{q_2}(0). 
\label{G0eq} \eea
\subsection{Exact scattering matrix}
We define the two-particle scattering matrix following \cite{Goorden07}; here we suppress the lead index. 
The two particle-scattering matrix is given by
\bea
S(E_1,E_2,E_3,E_4)\delta(E_1+E_2-E_3-E_4)=\la\psi^-_{E_1,E_2}|\psi^+_{E_3,E_4}\ra. \label{scattmat} 
\eea
The energy or momentum indices in the two-particle outgoing state indicate the energy or momentum of the two incident electrons.
After some rearrangement using the Lippman-Schwinger equation, we find
\bea
\la\psi^-_{E_1,E_2}|\psi^+_{E_3,E_4}\ra&=&\la\phi_{E_1,E_2}|\phi_{E_3,E_4}\ra+\la\phi_{E_1,E_2}|\f{1}{(E_3+E_4-\mathcal{H}_0+i \e)}\mathcal{V} |\psi^+_{E_3,E_4}\ra \nn \\
&+&\f{1}{(E_1+E_2-E_3-E_4+i \e)}\la\phi_{E_1,E_2}|\mathcal{V} |\psi^+_{E_3,E_4}\ra \nn \\ &=&\delta(E_1-E_3)\delta(E_2-E_4)-2\pi i \la\phi_{E_1,E_2}|\mathcal{V} |\psi^+_{E_3,E_4}\ra. 
\label{scattmat1} 
\eea
Finally we calculate the change of the two-particle scattering matrix due to the interaction,
\bea
\delta S(E_1,E_2,E_3,E_4)&=& -2\pi i \la\phi^{II}_{E_2}|\la\phi^{I}_{E_1}| \mathcal{V} |\psi^+_{E_3,E_4}\ra \nn \\
&=&-2\pi i \lambda \la\phi^{II}_{E_2}|0\ra \la\phi^{I}_{E_1}|0\ra \la 0,0|\psi^+_{E_3,E_4}\ra \nn \\
&=& -2\pi i \lambda \f{ \phi^*_{E_1,E_2}({\bf 0})\phi_{E_3,E_4}({\bf 0})}{1-\lambda K_{(E_3,E_4)}({\bf 0})}
\label{scattmat2} 
\eea
where we have used Eq.(\ref{psi0}) in the last line. In the weak coupling limit, i.e., $\lambda \to 0$, one gets back the corresponding expression of the two-particle scattering matrix of \cite{Goorden07}. Eq.(\ref{scattmat2}) is one main result of this paper. We emphasize that due to the interaction the two particles can exchange energy after scattering.
\subsection{Two-particle current}
 The current density in the conductor $I$ is given by the expectation value of the operator, $j^{I}_l = -i(c_l^\dg c_{l+1}-h.c.)$ in the two-electron scattering state $|\psi_\bk\ra=|\phi_\bk\ra +|S_\bk\ra$ (from Eq.(\ref{lipp})). The current in the incident state is given by 
\bea
\la \phi_\bk| j^{I}_l |\phi_\bk \ra =2 {\cal N_{I}}|t^I_{k_1}|^2\sin(k_1),
\eea
where ${\cal{N_{I}}}$ (a normaliasation factor) is the total number of sites in the conductor $I$. Similarly, one can find the current in the conductor $II$ in  the case $\lambda=0$, $\la \phi_\bk| j^{II}_l |\phi_\bk \ra =2 {\cal N_{II}} |t^{II}_{k_2}|^2\sin (k_2)$, where ${\cal{N_{II}}}$ is the total number of sites in the conductor $II$. The change in the current in the conductor $I$ due to scattering, $\delta j^I(k_1, k_2) =\la \psi_\bk |j^I_l|\psi_\bk \ra - \la \phi_\bk | j^I_l | \phi_\bk \ra$, gets contributions from two parts, namely, $j^I_S= \la S_\bk|j^I_l|S_\bk \ra$ and $j^I_C=\la S_\bk|j^I_l|\phi_\bk\ra +\la\phi_\bk|j^I_l |S_\bk\ra$.
\bea
j^I_S=2~{\rm Im}[\la S_{\bf k}|c^{\dg}_lc_{l+1}|S_{\bf k}\ra]
&=&2~{\rm Im}\Big[\f{\lambda^2|\psi^{+}_{\bf k}({\bf 0})|^2}{2\pi}\int_{-\pi}^{\pi}dq_1~ I_0(q_1)I^*_1(q_1)\Big] \label{autocorr} \\
{\rm with}~~I_s(q_1)&=&\f{1}{2\pi}\int_{-\pi}^{\pi}dq\f{\phi_{qq_1}({\bf 0})}{E_{\bf k}-E_{qq_1}-i\epsilon}\phi^{I*}_{q}(l+s)~~~{\rm with~s=0,1} \nn 
\eea
\bea
j^I_C&=&2~ {\rm Im}[\la \phi_{\bf k}|(c^{\dg}_lc_{l+1}-c^{\dg}_{l+1}c_{l})|S_{\bf k}\ra] \nn \\
&=&2~ {\rm Im}\Big[\f{\lambda\psi^{+}_{\bf k}({\bf 0})}{2\pi}\int^{\pi}_{-\pi}dq \Big( \f{\phi^{I*}_{k_1}(l)\phi^I_{q}(l+1)\phi^*_{qk_2}({\bf 0})}{E_{k_1}-E_{q}+i\epsilon}-\f{\phi^{I*}_{k_1}(l+1)\phi^I_{q}(l)\phi^*_{qk_2}({\bf 0})}{E_{k_1}-E_{q}+i\epsilon} \Big)\Big]
\label{crosscorr} 
\eea
If we switch off the on-site energy in the two interacting dot sites and take the hopping energy indentical to that of the leads, i.e., $\epsilon_I=\epsilon_{II}=0,~\g^I=\g^{II}=1$,  then we are able to  integrate the Eqs.(\ref{autocorr},\ref{crosscorr}) analytically and the total two-particle current change is given by
\bea
\delta j^I(k_1, k_2) =\f{{\rm Im}[K_{E_{\bk}}({\bf 0})]}{|1/\lambda- K_{E_\bk}(\b0)|^2}[{\rm sgn}(k_1)+{\rm sgn}(k_2)]
\eea
 Thus far we could not calculate Eqs.(\ref{autocorr},\ref{crosscorr}) analytically for arbitrary values of $\epsilon_I,~\epsilon_{II},\g^I,\g^{II} $, instead we evaluate them numerically \cite{Dhar08}. We find that the two-particle current change due to the interaction is smaller by a factor $\mathcal{N}_I$ than the incident current; this signifies that the probability of two-electron collision in conductor $I$ is order of $1/\mathcal{N}_I$. 
\subsection{Periodically varying on-site energy}
\begin{figure}
\begin{center}
\includegraphics[width=8.5cm]{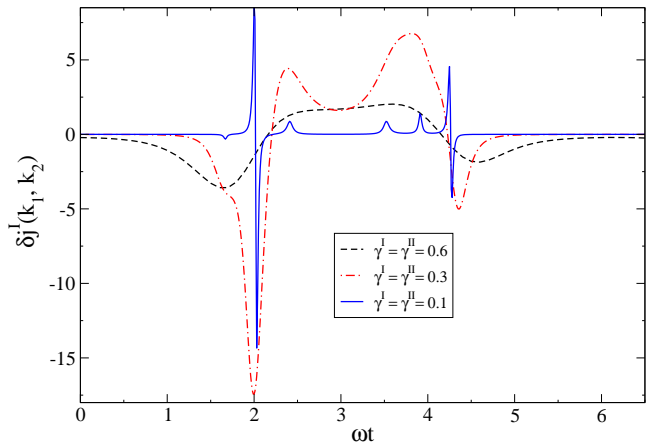}
\end{center}
\caption{(color online) Plot of the interaction induced two-particle current change in conductor $I$ with time $(t)$ for different tunneling strength. Here $k_1=k_2=1.35,~\phi=0.35,~ \lambda=0.8$ and $\omega=0.01$.}
\label{pumping2}
\end{figure}
In a recent experiment \cite{McClure07} the noise cross-correlation of two capacitively coupled QDs in the Coulomb blockade regime has been measured and the sign of this correlation has been found to change sign with tuning the on-site energy of the dot site by the gate voltage. Inspired by this experiment we now evaluate the two-particle current change for a periodically varying on-site energy of the two dots. This is like a two-electron quantum pump and we wish to study exactly how the strength of the interaction or the tunneling affect the current change averaged over a full cycle. We use,
\bea
\epsilon_{I}(t)=\cos(\omega t)~~{\rm and},~~~\epsilon_{II}(t)=\cos(\omega t+\phi)
\eea
In Fig.\ref{pumping2}, we plot the interaction induced two-particle current change $\delta j(k_1,k_2)$ with time for different values of the tunneling between the dot and the leads.
One can understand qualitatively that for which values of the on-site energy the sign of $\delta j^{I}(k_1,k_2)$ is changing. In the absence of the interaction $(\lambda=0)$, for a fixed incident energy $E_{k_1}(=-2\cos k_1)$, a single particle resonance occurs at the on-site energy of $\epsilon_I$, given by
\bea
\epsilon_I=E_{k_1}-{\g^I}^2(E_{k_1}\mp\sqrt{E^2_{k_1}-4})
\eea 
In the weak coupling limit $\g^I \to 0$ we expect $\epsilon_I=E_{k_1}$. Indeed we find that the change in sign of $\delta j^{I}(k_1,k_2)$ (like in Fig.\ref{pumping2} for the weak coupling case) occurs for the on-site energies correspond to the single particle resonance. 

\begin{figure}
\begin{center}
\includegraphics[width=8.5cm]{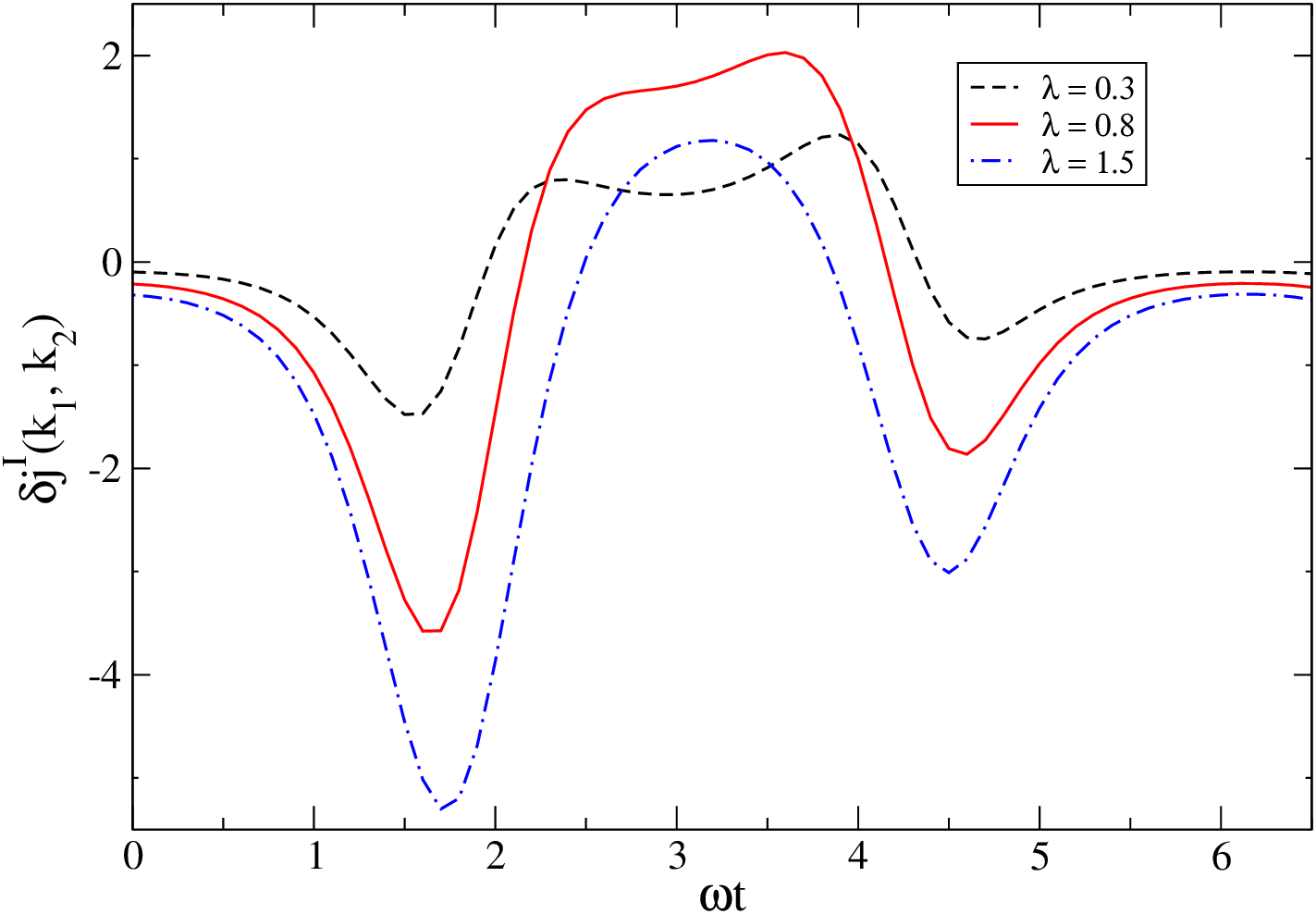}
\end{center}
\caption{(color online) Plot of the interaction induced two-particle current change in conductor $I$ with time $(t)$ for different interaction. Here $k_1=k_2=1.35,~\phi=0.35,~\g^I=\g^{II}=0.6 $ and $\omega=0.01$.}
\label{pumping3}
\end{figure}

Also we see in Fig.\ref{pumping3} that the average two-particle current ($\la \delta j^I(k_1, k_2) \ra$) depends on the interaction strength and can be positive or negative depending on the interaction. $\la \delta j^I(k_1, k_2) \ra$ crosses over from a positive to a negative value as the interaction strength increases. We define
\bea
\la \delta j^I(k_1, k_2) \ra=\f{1}{T}\int^{T}_0 dt j^I(k_1, k_2)(t)
\eea
Finally we write down formally the expressions (24-26) of Ref.\cite{Goorden07} for arbitrary  interaction.
\bea
\la \hat{n}^{I}_{R}\ra&=&\la \psi_{k_1,k_2}|c^{\dg}_{l}c_{l}|\psi_{k_1,k_2}\ra=\sum_m|\psi_{k_1,k_2}(l,m)|^2~,~~~~{\rm l>0,~k_1>0} \nn\\ 
\la \hat{n}^{I}_{R}\hat{n}^{II}_{R}\ra&=&\la \psi_{k_1,k_2}|c^{\dg}_lc_ld^{\dg}_{m}d_{m}|\psi_{k_1,k_2}\ra=|\psi_{k_1,k_2}(l,m)|^2~,~~~~{\rm l,m>0,~k_1,k_2>0}\nn \\
\la \delta \hat{n}^{I}_{R}\delta \hat{n}^{II}_{R}\ra&=&\la \hat{n}^{I}_{R}\hat{n}^{II}_{R}\ra-\la \hat{n}^{I}_{R}\ra \la\hat{n}^{II}_{R}\ra
\eea
As the two-particle scattering state, $\psi_{k_1,k_2}(l,m)$ is  known explicitly from Eqs.(\ref{scatt}-\ref{psi0}), one can calculate these correlations for any $\lambda$.
\section{A side-coupled interacting quantum dot acting as two-electron spin filter}
In the parallel conductors model one expects that the two electrons in the different conductors get entangled (orbital or pseudospin entanglement) due to the interaction. Here we study another mesoscopic system with the localized interaction which acts as a two-electron spin filter, i.e., the side-coupled interacting quantum dot filters out a two-electron spin-singlet state in the output lead from two-electron mixed input states in the input lead of the noninteracting quantum wire. Recently there is one similar study with the Anderson impurity model for the linear energy-momentum dispersion of the leads \cite{Inamura09}. 

\begin{figure}
\begin{center}
\includegraphics[width=6cm]{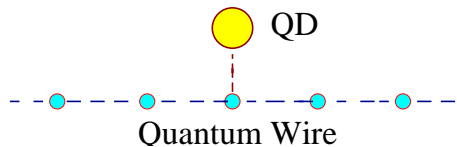}
\end{center}
\caption{(color online) A schematic description of a quantum wire with a side-coupled quantum dot (QD) modeled as an Anderson impurity.}
\label{side}
\end{figure}

We consider an interacting QD side-coupled to a perfect quantum wire (QW) [see Fig.\ref{side}] modeled by a single electron tight-binding Hamiltonian. The dot consists of a single, spin-degenerate energy level with an on-site Coulomb interaction between electrons. The main idea of our scheme is to prevent single-electron tunneling as well as current in the spin-triplet channel in the output lead. In that sense our program here matches with that of Oliver et al. \cite{Oliver02}. But we achieve these criteria through different mechanism and here, we don't need a three-port quantum dot geometry as well as leads acting as ``energy filters''.  We avert single-electron tunneling in the output lead by tuning the voltage gate attached to the QD. It occurs due to destructive interference of the single electron wave when the on-site energy of the dot-site is same as the energy of the incident electrons. It turns out that the above condition is also sufficient for complete destructive interference in the spin-triplet state. On the other hand in the presence of a finite Coulomb interaction in the dot, we get a finite current solely comprises of the two-electron spin-singlet state. Thus exchange interaction and quantum interference mediate to filter out the spin-singlet state of a two-electron mixed-state input of opposite spins. 
The full Hamiltonian $\mathcal{H}$ of the system consists of a noninteracting part $\mathcal{H}_0$ and an on-site Coulomb interaction part $\mathcal{V}$.
\bea 
\mathcal{H} &=& \mathcal{H}_0+\mathcal{V} \nn \\
\mathcal{H}_0&=& - \sum_{l=-\infty, \sigma=\ua,\da}^{\infty} (c_{l,\sigma}^\dg c_{l+1,\sigma} + h.c.)+\epsilon_{d} \sum_{\sigma=\ua,\da}n_{d \sigma}\nn \\&&-\gamma\sum_{\sigma=\ua,\da} (c_{0,\sigma}^\dg c_{d,\sigma} + h.c.) \nn \\
\mathcal{V}&=& U n_{d \ua} n_{d \da}~.\label{ham} 
\eea
where $n_{d,\sigma}=c^\dg_{d,\sigma}c_{d,\sigma}$ is the number operator in the dot for spin $\sigma$ and $\epsilon_{d}$ is the on-site dot energy.  $U$ is the strength of the on-site Coulomb energy in the dot site and $\g$ represents the tunneling strength between the quantum wire and the quantum dot. Again we set the hopping amplitude between sites on the quantum wire to unity.
\subsection{Scattering states}
For an electron coming from the left, the eigenstates $|\phi^{\sigma}_k \ra$ of $\mathcal{H}_0$ in the position basis $|l\ra$ are given by
\bea \phi^{\sigma}_k (l) &=& e^{ikl} ~+~ r_k e^{-ikl} ~~{\rm for}~~ l \le 0, \nn \\
&=& \frac{\gamma t_k}{\epsilon_{d}+2\cos k}~~~~~~{\rm on ~the ~dot~site}, \nn \\
&=& t_k e^{ikl} ~~~~~~~~~~~~~~~{\rm for}~~ l \ge 1, \eea
where $\phi^{\sigma}_k (l)=\la l|\phi^{\sigma}_k \ra$ and $0 < k \le \pi$ (with $\sigma=\ua, \da$).  The transmission, reflection amplitudes $t_k,~r_k$ are determined by solving the single electron Schr\"odinger equation; they are,
\bea
t_k=\f{2i (2\cos k+\epsilon_{d}) \sin k}{2i (2\cos k+\epsilon_{d}) \sin k + \g^2}~{\rm and}~ r_k=t_k-1~,
\label{trans1}
\eea
The incident energy of a single electron is $E_k = - 2 \cos k$. From the Eqs.(\ref{trans1}), we see that for a finite $\g$, the transmission amplitude vanishes at a finite incident energy, $E_k=\epsilon_d$, i.e., when the on-site dot energy is same as the energy of the incident electron. We can achieve this criterion by controlling the plunger gate acting on the QD. As before we calculate the single electron tunneling current  by taking expectation value of the current operator $j^{\sigma}_l=-i(c_{l, \sigma}^\dg c_{l+1,\sigma}-h.c.)$ in the single electron scattering state $|\phi^{\sigma}_k \ra$. Then, the single electron tunneling current in the output lead is given by $2 \mathcal{N}|t^{\sigma}_k|^2\sin k$, which vanishes at $E_k=\epsilon_d$. Here $\mathcal{N}$ is a normalisation factor depicting the total number of sites in the entire system. Now we consider that a spin-up and a spin-down electron incident in the input lead of the QW. The two-electron input state (with total $S_z=0$) is a mixture of spin-singlet and spin-triplet states whose spatial wave-functions are respectively symmetric and anti-symmetric. So the on-site Coulomb repulsion can cause scattering between two electrons in the spin-singlet channel but not in the spin-triplet channel. We find the current contribution from the spin-triplet channel in the output lead by taking expectation of $\sum_{\sigma}j^{\sigma}_l$ in the spin-triplet scattering state. If momentum of the two incident electrons are $(k_1,k_2)$, the current in the spin-triplet channel is given by,~ $2 {\cal N}(|t^{\ua}_{k_1}|^2\sin(k_1)+|t^{\da}_{k_2}|^2\sin(k_2))$.~  Then for vanishing current in the triplet channel we need to satisfy, $E_{k_1}=E_{k_2}=\epsilon_d$.  The last criterion also eliminates the possibility of both up or down spins (with total $S_z=\pm 1$) in the spin-triplet part of the input channel.

Now we calculate the effect of Coulomb interaction on the scattering of electrons in the spin-singlet channel. As there is no spin-flip interaction in the $\mathcal{H}$, we need to consider only the spatial part of the spin-singlet wave function. The scattering of two electrons in the spin-singlet channel can be studied using the LS formalism of Ref.\cite{Dhar08}. We consider here coherent electron transport at zero temperature. Incoming state of two electrons in the position basis is given by $\varphi_\bk(\bl)=\phi^{\ua}_{k_1}(l_1)\phi^{\da}_{k_2}(l_2)$ with ${\bf k}=(k_1,k_2)$ and ${\bf l}=(l_1,l_2)$. Then using the LS equation we find the two-electron scattering eigenstate $|\psi\ra$ of $\mathcal{H}$ as
\bea 
\psi_\bk(\bl) &=& \varphi_\bk(\bl) ~+~ U K_{E_\bk}(\bl)~ \psi_\bk ({\bf d}), 
\label{scattside} \\
{\rm where}~~ K_{E_\bk}(\bl) &=& \la \bl| G_0^+(E_\bk) |{\bf d} \ra~ {\rm and}~ \psi_\bk({\bf d}) ~=~ \f{\varphi_\bk({\bf d})}{1-U K_{E_\bk}({\bf d})}.  \nn \eea
 with ${\bf d} \equiv (d,d)$.  
Again the two-electron scattering eigenstate of $\mathcal{H}$ is completely given by Eq.(\ref{scattside}) with
\bea 
K_{E_\bk}(\bl)= \int_{-\pi}^\pi \int_{-\pi}^\pi \f{dq_1 dq_2}{(2 
\pi)^2} ~\f{1}{E_\bk - E_{\bf q} +i \e}\varphi_{\bf q}(\bl)\varphi^*_{\bf q}({\bf d}) . 
\label{G0eqside} 
\eea

\subsection{Spin-singlet current}
When the energy of the incident electrons is same as the on-site energy of the QD, there is no single electron tunneling as well as a zero current in the spin-triplet channel. So the current in the output lead is solely determined by the contribution from the spin-singlet channel. Now we determine the two-electron current density by the expectation value of the operator $\sum_{\sigma}j^{\sigma}_l$ in the scattering state $|\psi_\bk\ra=|\varphi_\bk\ra +|S_\bk\ra$ (from Eq.(\ref{scattside})). We calculate different parts of the current in the spin-singlet channel separately. The current in the incident state (with incident wave-vector $k_1,k_2$) is given by 
\bea
\la \varphi_\bk| \sum_{\sigma}j^{\sigma}_l |\varphi_\bk \ra =2 {\cal N}(|t^{\ua}_{k_1}|^2\sin(k_1)+|t^{\da}_{k_2}|^2\sin(k_2)),
\eea
which vanishes for $E_{k_1}=E_{k_2}=\epsilon_d$ identically.
\begin{figure}
\begin{center}
\includegraphics[width=8.5cm]{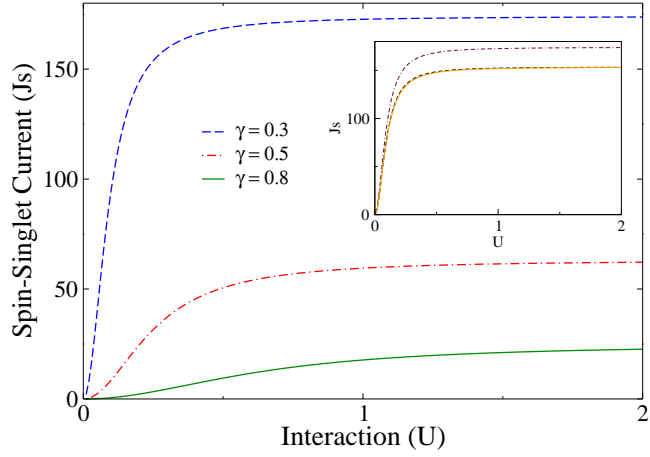}
\end{center}
\caption{(color online) Plot of the spin-singlet current arising from $Js=\sum_{\sigma}j^{\sigma}_s(k_0,k_0)/2$ vs interaction $(U)$ for $\epsilon_d=-0.2$ and different values of the coupling strength $\gamma$. Inset shows a plot of $j^{\sigma}_s(k_0,k_0)$ with $U$ for three different values of $\epsilon_d$ and a fixed $\gamma=0.3$. In the inset, the upper curve is for $\epsilon_d=-0.2$ and the lower two (almost overlapped) curves are for $\epsilon_d=-0.6,0.6$ respectively.}
\label{spincurrent}
\end{figure}
The change in the current due to the interaction, $\delta j^{\sigma}(k_1,k_2)=\la \psi_{\bk}|j_l^{\sigma}|\psi_{\bk}\ra-\la \varphi_{\bk}|j_l^{\sigma}|\varphi_{\bk}\ra=j^{\sigma}_s+j^{\sigma}_c$, with  $j^{\sigma}_s=\la S_k|j^{\sigma}_l|S_k\ra$ and $j^{\sigma}_c=\la S_k|j^{\sigma}_l|\varphi_k\ra+ \la\varphi_k|j^{\sigma}_l|S_k\ra$. For electron with $\ua$ spin, the current change solely from the scattered wave-function $|S_\bk\ra$ is given by
\bea
&&\la S_\bk|j^{\ua}_l|S_\bk \ra ~=~2 ~{\rm Im}\Big[ U^2|\psi_{\bf k}({\bf d})|^2~ \int \f{dq_1}{2\pi}|\phi^{\da}_{q_1} (d)|^2 \int \f{dq_2}{2\pi}\f{\phi^{\ua}_{q_2}(l+1)\phi^{\ua *}_{q_2}(d)}{E_{\bk}-E_{q_1q_2}+i\epsilon}\int \f{dq_3}{2\pi}\f{\phi^{\ua *}_{q_3}(l)\phi^{\ua }_{q_3}(d)}{E_{\bk}-E_{q_1q_3}-i\epsilon} \Big]
\label{auto}
\eea
This expression is similar in form of Eq.(\ref{autocorr}). Now $\psi_{\bk}({\bf d})$ is nonzero even for $k_1=k_2=\cos^{-1}(-\epsilon_d/2)=k_0$ and the integrand of Eq.(\ref{auto}) can not be said to be zero a priori. So we expect to have a finite contribution in $\delta j^{\sigma}(k_0,k_0)$ from this part. The other term in the current change comes from the overlap between $|S_\bk\ra$ and $|\varphi_{\bf k}\ra$, which is given by
\bea
&&\la S_\bk|j^{\ua}_l|\varphi_{\bf k}\ra+ \la\varphi_{\bf k}|j^{\ua}_l|S_\bk\ra \nn \\
&=&2 ~{\rm Im}\Big[U\psi_{\bf k}({\bf d})\phi^{\da *}_{k_2} (d) \Big\{\phi^{\ua *}_{k_1}(l)\int \f{dq_1}{2\pi}\f{\phi^{\ua }_{q_1}(l+1)\phi^{\ua *}_{q_1}(d)}{E_{k_1}-E_{q_1}+i\epsilon}-\phi^{\ua *}_{k_1}(l+1) \int \f{dq_1}{2\pi} \f{\phi^{\ua }_{q_1}(l)\phi^{\ua *}_{q_1}(d)}{E_{k_1}-E_{q_1}+i\epsilon}\Big\}\Big]~.
\label{cross}
\eea
The factors $\phi^{\ua *}_{k_1}(l)$ and $\phi^{\ua *}_{k_1}(l+1)$ in Eq.(\ref{cross}) vanish for $k_1=k_0$ in the output lead, i.e., $l>0$. So there is no contribution in $\delta j^{\sigma}(k_0,k_0)$ from the term in Eq.(\ref{cross}) if we evaluate current change in the output lead for electrons being incident from $l<0$. Ultimately we need to evaluate the integral in Eq.(\ref{auto}) to quantify the amount of spin-singlet pair generated in the output lead. As the parallel conductors model we determine it numerically for different values of the coupling strength $\gamma$ and the on-site dot energy $\epsilon_d$. We plot $Js(k_0,k_0)=\sum_{\sigma}j^{\sigma}_s(k_0,k_0)/2$ with interaction $U$ for different $\gamma$ in Fig.\ref{spincurrent}, which shows that the spin-singlet current increases with weaker coupling of the QD with the transport channel. This can be understood from the Eq.(\ref{auto}). $j^{\sigma}_s(k_0,k_0)$ depends on $|\psi_{\bf k_0}|^2$ which is inversely proportinal to $\gamma^4$. Occupation probability of the singlet pair  at the QD is higher for smaller $\gamma$; so the electrons scatter strongly with smaller $\gamma$ and $j^{\sigma}_s(k_0,k_0)$ increases. Fig.\ref{spincurrent} also shows that  $Js(k_0,k_0)$ saturates after some critical strength of interaction, $U_c$ which becomes smaller with decreasing $\gamma$. Here we should also clarify that one needs a finite coupling of the QD with the quantum wire to get a antiresonance in the single electron tunneling. We plot $Js(k_0,k_0)$ for three different values of $\epsilon_d$ in the inset of Fig.\ref{spincurrent}; we find that the magnitude of $Js(k_0,k_0)$ is same for a dot and an antidot on-site energy and the current increases with smaller dot energy. To check that the total current is same after scattering in both the input and the output leads, we evaluate $\delta j^{\sigma}(k_0,k_0)$ in the input lead also, i.e., $l<0$, in which case we need to evaluate both Eqs.(\ref{auto},\ref{cross}). We find total current is same in the input and the output leads within small numerical error. 
 
\section{conclusion}
We have calculated exactly the two-particle scattering state as well as the corresponding current in two interacting mesoscopic lattice models. In principle one needs to find a many-particle scattering state to study the out of equilibrium phenomena in these impurity models. But recently it has been shown in \cite{Dhar08, Nishino09, Inamura09} that many of the nonequilibrium quantities like the current-voltage characteristics have significant features in the two-particle current for weak interaction or low density of electrons. Though the many body effect drastically changes for strong interaction or higher density, for example, one expects to find an anti-Kondo resonance in the conductance of the side-coupled dot model in the presence of many electrons in the quantum wire.

We thank D. Sen and M. B{\"u}ttiker for many fruitful discussions and useful suggestions on the draft. The hospitality of Dept. of Theoretical Physics, Unversity of Geneva is gratefully acknowledged. 

$^\dagger$ Present address: Department of Physics, University of California-San Diego, La Jolla, California 92093-0319, USA

\end{document}